# *In Vivo* Wireless Sensors for Gut Microbiome Redox Monitoring

Spyridon Baltsavias, Will Van Treuren, Marcus J. Weber, Jayant Charthad, Sam Baker, Justin L. Sonnenburg, and Amin Arbabian

*Abstract*—A perturbed gut microbiome has recently been linked with multiple disease processes, yet researchers currently lack tools that can provide *in vivo*, quantitative, and real-time insight into these processes and associated host-microbe interactions. We propose an *in vivo* wireless implant for monitoring gastrointestinal tract redox states using oxidation-reduction potentials (ORP). The implant is powered and conveniently interrogated via ultrasonic waves. We engineer the sensor electronics, electrodes, and encapsulation materials for robustness *in vivo*, and integrate them into an implant that endures autoclave sterilization and measures ORP for 12 days implanted in the cecum of a live rat. The presented implant platform paves the way for long-term experimental testing of biological hypotheses, offering new opportunities for understanding gut redox pathophysiology mechanisms, and facilitating translation to disease diagnosis and treatment applications.

## I. Introduction

The human large intestine harbors trillions of microbes that interact with one another and the host. Identifying which microbial compositions ("microbiomes") promote health or disease is one of the grand challenges of the field. Different microbiomes produce different chemical environments in the intestine [1], [2] but how these microbiome-mediated chemical environments impact host physiology and health status are unclear. An emerging hypothesis is that disease-associated microbiomes share a common feature: an altered chemical landscape in the gut with a higher level of environmental oxidation [3]–[7].

Evidence for this hypothesis comes from multiple different systems. First, *Salmonella* and *Citrobacter* pathogens encourage inflammation in the host to promote an oxidative environment. This oxidative environment contains more terminal electron acceptors for direct use in respiration [8], [9] and produces new carbon substrates that are more readily catabolized by these pathogens [10], [11]. The oxidant-supported outgrowth of these aerotolerant pathogens potentiates inflammation, toxic metabolite production and impairs barrier function [5]; if the host immune response cannot control the pathogens, chronic inflammation ensues and positive feedback prevents recolonization with a healthy microbiome [12]. Second, depletion of butyrate production (along with other short-chain fatty acids [SCFAs]) in the gut is associated with expansion of pathogens and dysbiosis [13], [14]. The mechanisms are complex, but oxidant balance plays a key role: without colonic butyrate, colonocytes reduce beta-oxidation, leading to increased aerobicity in the gut [15], [16]. Finally, soil and seafloor microbial communities are structured by gradients in environmental oxidation state and the availability of terminal electron acceptors and donors [17]–[19]. Disturbances in the redox balance of these environments can impair function [20] in an analogous manner to redox balance in the gut.

Unfortunately, testing this oxidation hypothesis directly is currently impossible because no *in vivo*, longitudinal, oxidation-measuring sensor exists for research animals (e.g. mice and rats). Battery-powered capsules for human ingestion have been demonstrated for applications such as gas (e.g. oxygen, $CO_2$), pH, bioluminescence sensing and more [21]–[23]. However, capsule size is limiting for use in model organisms where biological hypotheses can be directly and conveniently tested (e.g. through dietary change, antibiotic treatment, or genetic alteration of host or microbe). Ingestible capsules also transit the gut rapidly, typically within 24 to 48 hours in humans [24], [25], prohibiting location-specific and longitudinal tracking of changes in gut microbiome states which occur over longer time frames [26], [27]. We propose an implantable device as a platform for extended and on-demand gut sensing in small animals. In this paper, we design such an implant sensor using oxidation-reduction potentials (ORP) as a measurement modality. ORP is an integrated measure of an environment's propensity to lose or gain electrons and has been used to explain microbial energetics and ecology primarily in environmental contexts [17], [28]. As outlined above,

This work has been submitted to the IEEE for possible publication. Copyright may be transferred without notice, after which this version may no longer be accessible. Supplementary information provided in the appendix.

This work was supported by: the DARPA Young Faculty Award (D14AP00043), the NSF CAREER Award (ECCS-1454107), the NSF Graduate Research Fellowships Program (DGE-114747), NIH NIBIB (R01EB025867), and the Stanford Bio-X Interdisciplinary Initiatives Seed Grants Program (IIP) R9113JSAA. The content is solely the responsibility of the authors and does not necessarily represent the official views of the National Institutes of Health.

S. Baltsavias, M. J. Weber, J. Charthad, and A. Arbabian are with the Electrical Engineering Department, Stanford University, Stanford, CA 94305 USA (contact email: sbaltsav@stanford). W. Van Treuren and J. L. Sonnenburg are with the Department of Microbiology and Immunology, Stanford University, Stanford, CA, 94305, USA. S. Baker is with the Department of Comparative Medicine, Stanford University, Stanford, CA 94305, USA.

evidence suggests diet and antibiotic-induced dysbiosis changes the energetic landscape of the gut, substantially increasing the concentration of high-potential terminal electron acceptors (oxidants) like oxygen, reactive oxygen and reactive nitrogen species (ROS/RNS) [5], [13]. Recent work has applied ORP sensing to fecal samples from mice, demonstrating significant ORP changes due to antibiotics [6], but the relevance of fecal ORP to gut physiological conditions is unclear. An internal real-time ORP measurement using an implantable sensor has not been demonstrated. Towards this goal, we design a prototype ORP sensor comprising a custom CMOS IC with ultrasonic, wireless power up and telemetry, robust biocompatible packaging for longevity in the harsh gut environment, and solid-state ORP electrodes. In the following sections, we outline the underlying theory of operation, the design and *in vitro* characterization of the sensor components, and the interaction of electronics and electrodes in biological media. We conclude this paper with a proof-of-concept 12-day *in vivo* measurement of the fully integrated and encapsulated sensor in a rat.

## II. PRINCIPLE OF OPERATION

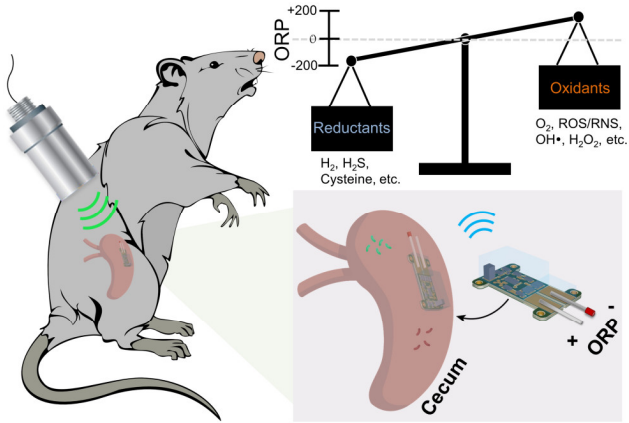

Fig. 1: Conceptual description of redox sensor operation. The sensor is wirelessly powered and measures the balance of oxidants and reductants (the ORP) produced by the interplay of host, microbe, and diet.

The principle of operation of the proposed redox sensor is shown in Fig. 1. The sensor is surgically implanted in the cecum of a rodent; while multiple sites for implantation are possible, the cecum contains the majority of bacteria in the rodent by mass, and is thus most attractive for understanding host-microbe interactions [29]. An external transducer placed in contact with the body sends power via ultrasonic waves to the sensor on demand. The sensor harvests the received power, measures the ORP, and transmits ultrasound back representing the measured data.

The ultrasonic link used for power delivery and data transmission relies on far-field radiative power transfer, allowing the external transducer and implant transceiver to be designed separately [30], [31]. In this work we use a commercial single-element ultrasound probe for the external device and focus on the design of the implant. We utilize ultrasonic links because they allow high acoustic-electrical efficiency with mm-sized transducers and exhibit lower tissue attenuation (~0.5-1 dB/cm/MHz) than other methods such as RF/inductive powering [32], [33]. These advantages allow for reliable and safe operation in rodents (well below the FDA diagnostic ultrasound intensity limit, 7.2 mW/mm²), while enabling scalability to larger animals and future translational studies.

ORP is a potential difference developed between the sensor electrodes as a result of spontaneous redox reactions of luminal chemical species. ORP electrodes consist of an inert metal indicator/working electrode, in this case a platinum wire, and a reference electrode. In a complex chemical environment such as the gut, ORP is approximated as the weighted sum of all potentials of redox couples present [28]. The resulting potential difference (working electrode minus reference) for $m$ redox couples can be expressed as, [28],

$$E(V) = \sum_{i=1}^{m} \frac{|i_i^0|}{\sum |i_i^0|} [Eh_i^0 - \frac{RT}{n_i F} ln \frac{(Red_i)}{(Ox_i)(H^+)^a}], \quad (1)$$

where $i_i^0$ is the electron exchange current for each couple, $Eh_i^0$ is the individual couple redox potential at standard state, $R$ is the gas constant, $T$ is absolute temperature, $n_i$ is the number of electrons exchanged for each reaction, $F$ is the Faraday constant, $(Red_i)$, $(Ox_i)$, $(H^+)$ are the activities of reductants, oxidants and hydrogen ions respectively, and $a$ is the number of protons exchanged. Thus, an ORP measurement captures a 'snapshot' of the oxidation-reduction profile of an environment that is predominantly shaped by the most abundant redox couples. In a healthy rat cecum, ORP with respect to a silver/silver chloride (Ag/AgCl) reference electrode has been reported in the range of -500 mV to -100 mV[1] [34]–[36]. The more negative ORP is (the lower the working electrode potential), the greater the tendency of the chemical species in the medium to gain electrons (become reduced); therefore this range suggests the cecum is a strongly reducing environment, which is consistent with the lack of high-potential oxidants under normal conditions. While no electrode is truly inert or unbiased, the use of platinum in this sensor allows us to broadly capture the chemical species contributing to ORP in the gut, including oxidants like $O_2$ and ROS/RNS [17].

---

[1] Note that Wostmann and Bruckner-Kardoss [35] and Schröder et al. [36] report ORP (*Eh*) with respect to the standard hydrogen electrode, resulting in an approx. +200mV offset from measurements with an Ag/AgCl reference, such as the ones we report in this work. They also normalize to pH 7; while ORP depends on pH, the correction factor depends on the specific oxidation-reduction system [34]. Since the cecum is an unknown system, we did not apply a correction factor in our measurements.

## III. METHODS AND RESULTS

The implant sensor consists of three main components (Fig. 2a): electrodes for measuring ORP, encapsulation materials, and electronics including an ultrasonic piezoelectric transceiver (piezo) and an integrated circuit (IC) for ORP readout and wireless telemetry.

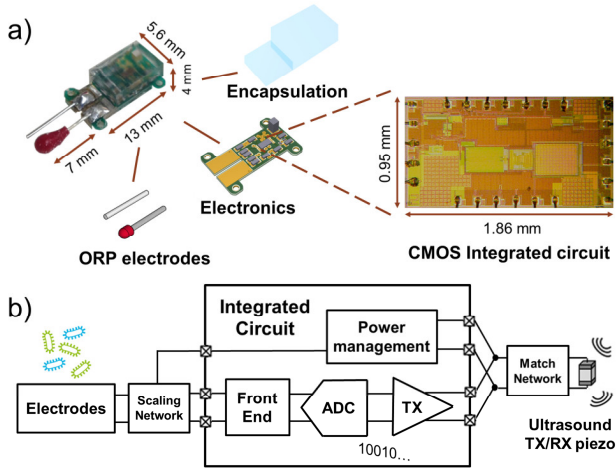

Fig. 2: a) Photograph of the designed and fabricated redox implant, and breakdown of its construction. b) Simplified system block diagram.

### A. Electrodes

An ideal reference electrode maintains a stable potential despite externally varying chemical conditions. Commercial ORP systems typically employ an Ag/AgCl reference electrode with a large volume of liquid electrolyte - a 'tank' (e.g. a Sigma-Aldrich Ag/AgCl electrode has 1-5 ml of 3 M KCl). This excess volume of $Cl^-$ ions nearly eliminates loss due to leakage across environmentally varying $Cl^-$ concentration gradients, ensuring a constant reference potential [37], [38]. Unfortunately, this type of electrode is challenging to adapt to the requirements of our sensor, due to the limited volume available. An alternative solution suited to size constraints is a solid-state reference electrode. Solid-state electrodes typically consist of an Ag wire covered by AgCl coating encapsulated in a polymer impregnated with KCl. They are designed to allow minimal ionic flow and prevent electrolyte leakage [37], [38]. Here we used solid-state electrodes (Refex Ltd, referred to as 'Refex') made of a silver wire, the tip of which is chloridised and covered with a KCl-doped vinyl ester resin [39]. In all tests we used a platinum (purity 99.99%) wire as the working electrode, with both platinum and Refex wire having diameter of 0.38 mm and length 7 mm.

To assess accuracy we compared ORP measured by Refex electrodes to those of a Sigma-Aldrich standard reference double junction Ag/AgCl electrode using a high impedance SympHony SB70P pH meter in ORP mode (Fig. 3a). Since the complete sensor must undergo sterilization before *in vivo* implantation, the Refex electrodes were sterilized with a standard autoclave cycle (121 °C for 15 minutes in 2.8 M KCl) prior to use. After a 12-hour rehydration period in 2.8 M KCl, Refex and platinum electrode pairs were placed in standard ORP calibration solutions (Light's solution, Zobell's solution, and modified Zobell's). Between solution immersions, electrodes were wiped dry to avoid cross-contamination. The developed voltage was recorded for 2 minutes and the average was taken. The Refex electrodes withstand sterilization and measure the ORP to within 15 mV of the commercial reference system with < 2% deviation in slope.

Next, we tested the dependence of the measured ORP on $[Cl^-]$ for Refex electrodes to determine how sensitive they would be to cecal changes in $[Cl^-]$ (Fig. 3b). An Ag/AgCl electrode with an infinite 'tank' should show no dependence on $[Cl^-]$ [40], while a Ag/AgCl electrode without any electrolyte volume (pseudo-reference) has an ideal response of approximately 59 mV change per decade change in $[Cl^-]$ based on the Nernst equation at room temperature. In this experiment, true Ag/AgCl gel-based reference electrodes (REDOX ORP-14, PCE Instruments) were immersed in constantly stirred KCl solutions with either a Refex or an Ag/AgCl pseudo-reference as counter electrode (see supplemental note A "Materials", and Fig. 3c). We performed multiple iterations of this experiment in short and longer time scales and measured the potential difference using a high impedance ORP meter (PCE-228-R, PCE Instruments). Minute-long immersions of Refex (after 12 hours of hydration in 2.8 M KCl) showed the electrodes maintain a stable potential over two decades of $[Cl^-]$ change. Over longer time scales (24 hours), while the pseudo-reference electrodes show a Nernstian response as expected, the Refex show a sub-Nernstian response, with an average slope for two electrodes of 11 to 14 mV/decade. We repeated this experiment with four Refex cycled through mixtures of 50% PBS 1X and 50% KCl 0.01 to 1 M (the total $[Cl^-]$ concentrations were calculated based on standard PBS values; see supp. section A). Here the Refex electrodes show a more complex behaviour, with chloride concentrations near 0.1 M resulting in comparable values but concentrations near 1 M deviating to higher values than the previous experiment. Human intestinal concentrations of $[Cl^-]$ are described over a range of physiological concentrations and appear to range between 50-150 mM [41]–[44]. Thus, assuming the response evident in 24-hour KCl incubations (Fig. 3b), ORP readings from Refex electrodes would vary < 7 mV due to physiological changes in chloride concentrations. For experiments spanning more than 24 hours, it is expected the Refex electrodes will approach a Nernstian response similar to a pseudo-reference Ag/AgCl due to loss of chloride ions in the polymeric matrix. This would result in a worst-case variation of approx. 30 mV in the gut. Other sources of variance in biological ORP readings are substantially greater than 7-30 mV suggesting Refex electrodes are well-suited for capturing the relevant redox biology occurring in the cecum[2]. The accuracy over time of these solid-state electrodes could be improved with a thicker layer of KCl-doped resin with the trade-off of longer hydration time required [45].

### B. Electronics

The goal of the electronics is to provide reliable sampling of ORP as well as wireless power up and communication capabilities. Wireless sensor implants often use analog circuitry and passive telemetry schemes (e.g. backscattering) [46], [47]; while their development and operation are simpler allowing for miniaturization, they lack digital precision and can be susceptible to signal loss when operating in attenuative tissue. Conversely, we use a custom-fabricated chip (TSMC CMOS 180nm HV) with an active uplink and 10-bit analog-to-digital conversion to maintain robustness and precision over long-term *in vivo* operation; the design of the IC has been detailed in previous work [32].

A simplified block diagram of the electronics is shown in Fig. 2b. The IC primarily comprises a power management unit (PMU), an analog front-end (AFE), a 10-bit Analog to Digital Converter (ADC) and a digital data transmitter (TX). The power consumption breakdown is shown in supplemental Fig. S1. The IC is wirelessly powered via ultrasonic waves, captured by the designed piezo, which initially serves as the acoustic to electrical power transducer. The system is designed to operate in bursts, allowing for on-demand sampling of the ORP at desired intervals. The onset of incoming ultrasound is detected using the PMU, and the electrical power from the piezo is rectified and stored in an off-chip storage capacitor. After an initialization phase, the AFE samples the ORP voltage input generated by the two electrodes. The AFE has been modified using a scaling network to sample ORP in the range of -500 mV to 300 mV, which covers the expected range in the gut [34], [35] and allows measurement of common ORP calibration standards. Since the AFE is originally designed for general purpose voltage sampling, we considered its input resistance and leakage current (Fig. S2) and their effect on our measurement accuracy, which we discuss in section "Tests in fecal media and electronics input resistance". After the AFE samples an input voltage, the sample is digitized by the ADC with sub-mV resolution, encoded using on-off keying (OOK) modulation and transmitted by a power amplifier (PA) in the TX block. OOK is used for simplicity of implementation and for energy savings, as the PA is turned off for transmission of bit 0. The modulated data is sent through the piezo, finally serving as an electrical to acoustic power transducer. The implant then powers off and waits for the next ultrasound power burst. To enable high temporal resolution, the power burst period and thus the sampling period is set to 1 ms for all measurements in this paper; since this parameter is externally controlled, it can be increased if needed to trade off measurement resolution with reduced average power dissipation.

The piezo utilized in this implant is a 1.1 x 1.1 x 1.8 mm³ PZT5A cube. Note that the same piezo is used for both uplink and power up through time domain multiplexing, saving area. We have previously extensively investigated the design of piezo transceivers for many sensing applications [32], [48]–[50]. In this application, the frequency range of operation for the power up as well as wireless data communication is 790-840 kHz. At these frequencies, the link loss through tissue is < 1 dB/cm [33]. The dimensions and material of the piezo are chosen to place this frequency range in the piezo inductive region [48]. In this region the piezo has a positive reactance, allowing us to use a series capacitor with a small area footprint to optimize the impedance. The desired piezo resistance was 10 to 30 kΩ in this work, to allow for both > 40% efficient power up (see supp. section B "Ultrasound transmission and data recovery" for more details) and as well as approximately 80 μW to 230 μW output power from the IC power amplifier [32]. Power levels in the 100 μW range have been shown to be sufficient for transmission through 12 cm of tissue phantom, resulting in high signal-to-noise ratio (SNR) at the external receiver [32]. Here, the transmission distance is much lower (~1 cm maximum in the body), allowing for margin in case of misalignment and other unexpected sources of loss.

*In vitro* electronics characterization was first performed to characterize the wireless ultrasonic powering and data communication at a controlled depth, as well as the correct sampling and digitization of input voltages (Fig. 3d, 3e). The setup (Fig. 3f; details in supp. section B) uses a custom acrylic-constructed tank filled up to 1 cm from the tank bottom with mineral oil; mineral oil is used to model the acoustic impedance of soft tissue [51] and 1 cm is chosen based on the maximum expected implant depth in tissue. An implant before encapsulation, consisting of the IC, piezo and other surface-mount components, was placed in the measurement tank bottom. Soldered wires at the AFE input terminals, extending 5 cm, allowed us to interface with standard ORP electrodes external to the tank. Tests were performed by immersing the electrodes in Zobell's and Quinhydrone (saturated in pH 7 buffer) standard solutions; to achieve a negative ORP the electrode terminals were switched and for a zero-voltage calibration point the leads were shorted together.

---

[2] Older literature [34]–[36] reported ORP measurements of > 10 minutes in length, with > 100 mV variations. Newer literature has reported the mean of 3-minute readings with > 100 mV standard deviations [6].

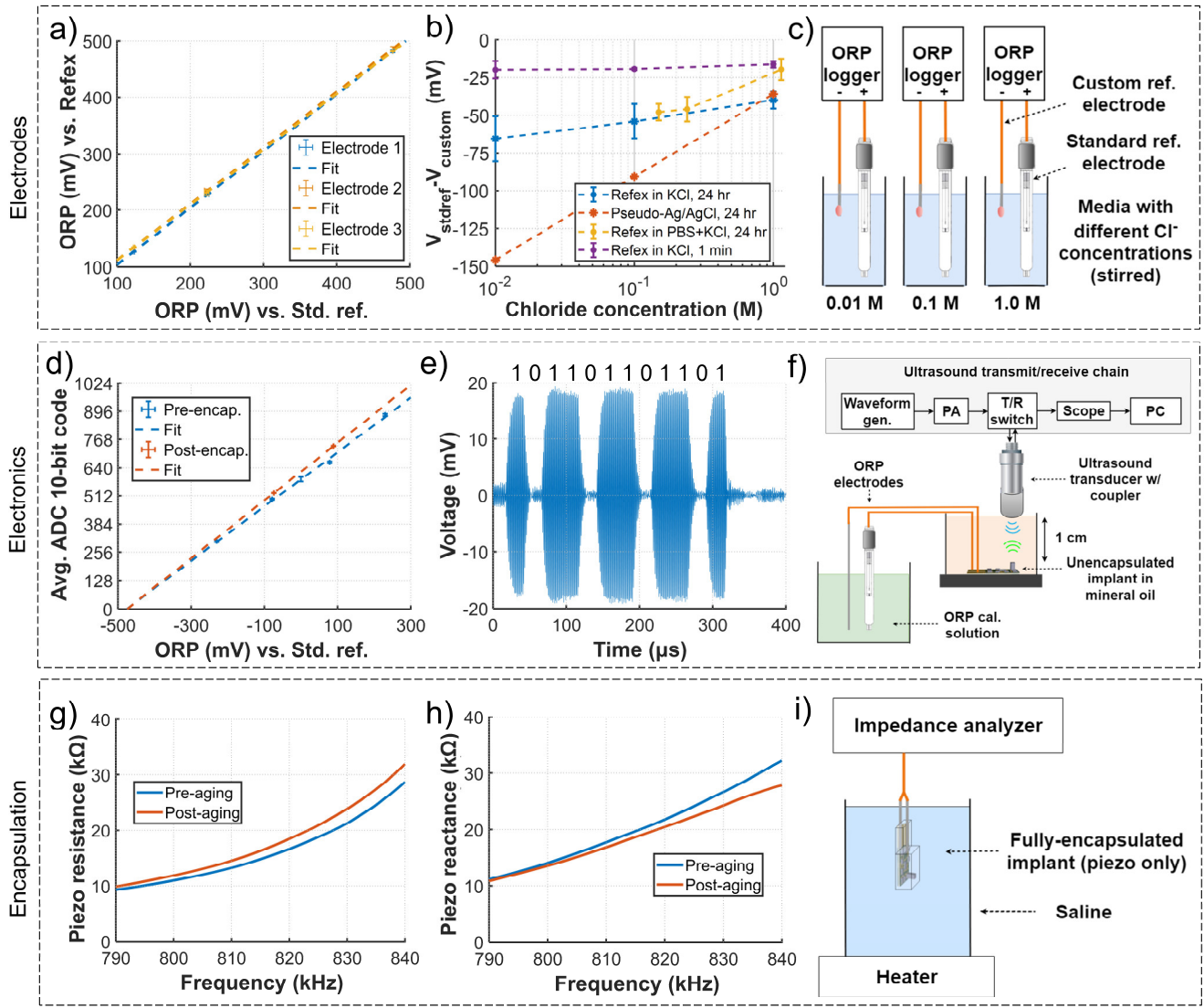

Fig. 3: *In vitro* tests of sensor components. a) Custom electrodes accurately measure standard media vs. a standard Ag/AgCl reference. b) Custom electrodes are resistant to external [Cl$^-$] changes for short term exposure. c) Experiment setup for subfigure b. d) The IC measures and successfully transmits ORP wirelessly before and after encapsulation and packaging. e) Example of OOK data wirelessly recorded from the sensor *in vitro*. f) Measurement setup for subfigures d (pre-encapsulation) and e. g), h) Pre and post-aging piezo resistance and reactance profile respectively indicate the encapsulation is robust to aging. i) Measurement setup for impedance characterization. Note: all error bars shown represent the standard deviation.

A focused single-element external ultrasound transducer was used to transmit ultrasound power to the implant piezo; the same transducer captured ultrasound signals from the piezo. The average output sensor code throughout 2 minutes of measurement was compared to the corresponding reading from a high-impedance ORP meter connected to the same electrodes. Fig. 3d (in blue) shows the measured ORP, indicating that the sensor backend output is linear ($R^2 = 0.9975$) to ORP measured with a commercial meter. An example of received sensor data (10-bit message representing measured ORP, preceded by a '10' preamble) with this setup is shown in Fig. 3e validating the wireless performance; the received SNR is ~30 dB enabling reliable data interpretation. Note that data reported in Figs. 3d, 3e, 5d, and 5e, are from the same sensor tracked throughout the fabrication and testing process.

To verify the insensitivity of the electronics to electrode properties and packaging, the implant board - after connection to a pair of custom electrodes and complete encapsulation (described in the "Fabrication and encapsulation" section) - was placed in two calibration solutions: Quinhydrone in pH 7 and pH 12 buffer. The setup used for this measurement (Fig. S3) was similar. The external transducer for powering and interrogating the sensor was immersed in the ORP calibration medium of interest at a distance of 1 cm. The resulting post-packaging measurement (in red) is overlaid with the pre-packaging measurement in Fig. 3c. The pre-encapsulation measurement uses a pair of standard electrodes, whereas the post-encapsulation sensor uses the custom electrodes.

The curves are very similar, showing that the sensor maintains linearity after packaging. A 35-code difference in offset and 1% difference in slope are attributed to the custom electrodes used for the specific sensor. The post-packaging measurement further serves as a calibration curve, used in subsequent *in vivo* tests to interpret sensor outputs in unknown media (see "*In vivo* demonstration" section).

### C. Fabrication and encapsulation

It is important for an *in vivo* implant to be easily manufacturable for large-n rodent studies, as well as water impermeable and biocompatible for long-term operation in the body. To minimize local inflammation response and prevent infection, the implant should also able to withstand sterilization procedures. Additional constraints in our case include the need for the encapsulation material to be compatible with ultrasound transmission, and with surgical suturing techniques for fixing the implant position in the rodent cecum. The fabrication strategy adopted to meet these constraints is as follows.

First, the sensor electrical components (IC, capacitors, and piezo) are assembled on a thin (200 μm) FR4 Printed Circuit Board (PCB). The thickness was chosen so as to limit the device volume and provide some flexibility while retaining structural integrity for component placement and bonding. For ease of fabrication, only the top side of the PCB is populated; for further miniaturization, both sides can be utilized. The implant has four PCB via-holes that extend from the main body of the board (see Fig. 2a), which we refer to as 'wings'. The via-holes on the wings accommodate a size 4:0 suture needle and allow attachment to the cecal epithelium. After assembly, the implants are tested as described in the "Electronics" section to verify functionality. The unencapsulated sensors are then cleaned via immersion in acetone, methanol, and isopropanol.

To provide water impermeability and biocompatibility, the implants are encapsulated in layers of Polydimethylsiloxane (PDMS) and Parylene C [52], [53]. PDMS is also advantageous because its acoustic impedance matches that of tissue [54]. A custom-designed aluminum mold holds the implant during PDMS curing, which lasts 48 hours (room temperature). The mold shape allows for complete encapsulation of the main body of the implant while leaving the electrode pads as well as the wings uncovered. The electrodes are attached with conductive epoxy, leaving the sensing tips protruding from the sensor body. An extra protective layer of nonconductive medical epoxy is then added on top of the electrode pads for biocompatibility and cured at 65 °C for an hour. Finally, the implants are coated with 6 μm of Parylene C. Since the Parylene C covers the sensor electrode tips, it is carefully removed to allow contact of 2 mm of the electrodes with the medium. Before *in vivo* implantation, the implants are tested to produce the final calibration curve as mentioned previously. Implants are sterilized by autoclaving while submerged in 2.8 M KCl and left to rehydrate in the same sterile solution. The final post-encapsulation implant measures 20 x 5.6 x 4 mm³ (Fig. 2a), allowing it to be implanted in a rat cecum with ease.

After validating the robustness of the electrodes and electronics in the previous sections, we separately tested the packaging robustness by fabricating a simplified sensor consisting of the piezo element only. Measurements of the piezo impedance and its changes through a polymer aging process were used as a test; major changes in the resistance and reactance values (e.g. indicative of a short) were interpreted as package failure and water ingress affecting the piezo component. To account for the presence of the electrode connections in the actual implant, the piezo terminals were internally wired to the implant electrode pads, which were connected to the impedance analyzer. Encapsulation and autoclave-sterilization were performed as described above and then an initial reference measurement of the piezo impedance was taken. Subsequently, the packaged 'mock' implant was immersed in a heated solution of saline (0.9%) for 24 hours to induce accelerated packaging aging and simulate the effects of *in vivo* implantation [55]. The aging solution temperature was 73 °C, giving a 12x acceleration [55] for total aging equivalent to a 12-day *in vivo* implantation. Afterward, a final measurement of impedance was taken and compared to the initial reference point. The results (Figs. 3g, 3h) indicate that the encapsulated piezo was negligibly affected by the aging procedure; the discrepancy is < 13% in the frequency region of interest, likely caused by movement in connection wires in the measurement setup (shown in Fig. 3i). This suggests that the encapsulation materials can withstand at least 12 days of *in vivo* operation, protecting the implant electronics from degradation.

### D. Tests in fecal media and electronics input resistance

In the previous sections we report experiments for verification of the chip and electrode functionality in standard media. Biological media like the contents of the gut lumen are significantly more complex, containing digesta, sloughed epithelial cells, immune cells, and hundreds of bacterial species [56]. Because of this chemical complexity, the number of chemical species contributing to the ORP is likely significantly higher. ORP measurements have been used in soil science for several decades to predict environmental traits of relevance for agriculture and groundwater measurement. Two phenomena encountered in soil measurements of ORP are the dependence of measured ORP on measuring-electronics impedance [17], [57], and the fouling/poisoning of the electrodes by biotic or abiotic chemical factors. Here, we investigated whether these phenomena also appear in biological fecal media and how they might affect our *in vivo* measurements in rodent ceca.

To test the effect of impedance on ORP, we anaerobically incubated 3 Refex and polished platinum electrode pairs and one standard reference and polished platinum pair for 12 days in a fecal slurry (50-90% feces, 50-10% PBS V/V in an anaerobic chamber with atmosphere $N_2:CO_2:H_2$ 80:15:5). Measurements with custom electrode pairs were started with a direct connection to a 1 T$\Omega$ ORP meter, (t = 0 to 0.2 hours) and then connected in parallel with a 17 M$\Omega$ resistor. The standard reference electrode pair was maintained at 1 T$\Omega$ as ground truth. At high impedance (t = 0) the Refex electrodes start within 20 mV of the standard reference (Fig. 4a inset). Upon connection to 17 M$\Omega$ resistance loading (setup shown in Fig. 4b), the ORP changes and over 12 days deviates ~300 mV from the ground truth, with a slow drift towards 0 (Fig. 4a); based on this result we verified that ORP measurements in feces depend on the impedance of the measuring equipment.

To determine whether the 12-day fecal exposure caused electrode fouling and permanent damage, especially under constant current draw due to low measurement resistance, we subsequently returned the custom electrodes to high impedance (1 T$\Omega$) by removing the 17 M$\Omega$ resistor (Fig. 4a, at t = 264 hours). Within 22 hours, the custom electrodes' ORP recovered to within 30 mV of the standard reference value (estimated asymptote using curve fitting; see supplementary Fig. S4). This indicates the electrode potential is, to first order, minimally affected by biofouling and extended electrical loading. Note that these results may be confounded by the custom electrode drift and offset due to medium [Cl⁻] as discussed in the electrodes section.

When the implant is immersed in the gut lumen, the electronics interact with electrodes both when they are off and when they are on (wirelessly powered). The front-end resistance was therefore measured over the entire input range in both off and on modes, using a sensitive current meter (Keysight B2962A). The worst case off-mode leakage was found to be 30 nA for -0.5 V inputs, and 3 nA in on-mode for the same input (supp. Fig. S2). This corresponds to approximately 17 M$\Omega$ impedance in off-mode and 170 M$\Omega$ in on-mode. Therefore, the experiment results in Fig. 4a ("chip off") also illustrate the worst-case effect of the implant electronics on ORP if the implant was powered off for 12 days *in vivo*. Note that the effect of the electronics input capacitance was neglected as a minor contributor to the overall response time constant.

Having established that the electrodes can maintain correct ORP with return to high impedance, we asked whether the implant on-mode resistance is high enough during wireless measurements to allow sufficient measurement accuracy. Repeating the above experiment in a different fecal matrix, for a Refex and Pt electrode pair we studied the transition from off mode to on mode, and then to an unloaded measurement (1 T$\Omega$). We similarly conservatively modeled the chip using resistors of 170 M$\Omega$ for on-mode and 17 M$\Omega$ for off-mode. We repeated the resistance switch sequence 3 times within an 8-day period (Fig. 4c). Here the on-mode settled value was consistently ~50 mV different from the unloaded value, indicating our implant would measure a higher ORP than a 1 T$\Omega$ instrument in this medium. The unloaded ORP itself was higher than that measured with a standard reference electrode and a different Pt at 1 T$\Omega$, by ~70 mV in 1 hour and ~25 mV in 22 hours (Fig. S5), which could be attributed to Refex electrode offset, differences in the individual fecal media in each of the containers, and changes to individual platinum electrodes.

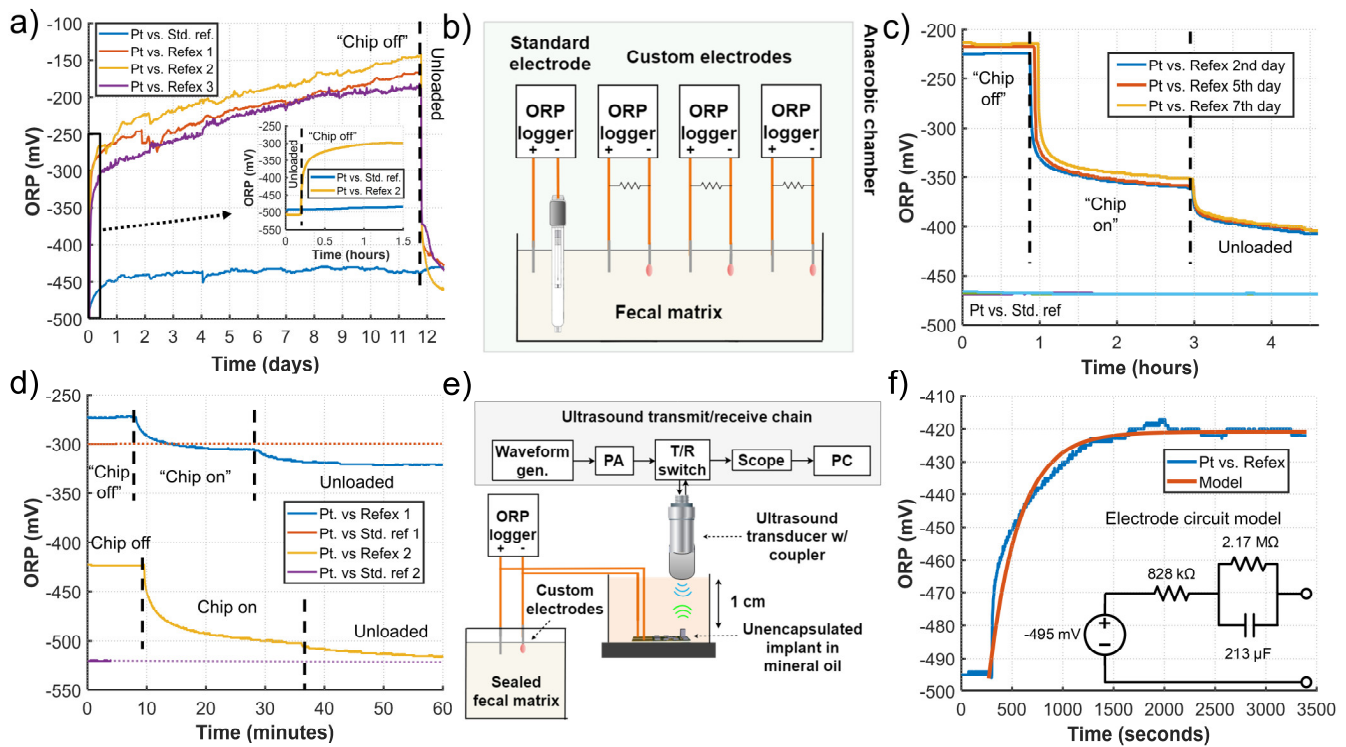

Fig. 4: a) Low measurement resistance (17 M$\Omega$) causes higher custom electrode ORP, but the ORP recovers when the resistance is disconnected, despite 12 days of fecal medium exposure. b) Measurement setup for subfig. a. c) Transitions in measurement resistance from 17 M$\Omega$ (electronics off mode) to 170 M$\Omega$ (on mode) consistently result in an ORP closer to the unloaded measurement. d) Additional examples of short-time scale transitions in different fecal media. Top: loading using resistors. Bottom: using an actual IC connected to electrodes. e) Measurement setup for subfig. d, bottom. f) The transitions in measurement resistance can be modeled using a circuit model similar to the Randle circuit.

We performed multiple repetitions of this experiment with short vs. long exposure of electrodes in fecal media as well as in stirred and unstirred conditions. An example of two short time scale experiments is shown in Fig. 4d. Here, similar behaviour was observed but with lower errors, with a modelled IC on-mode being ~20 mV different than the unloaded value (Fig. 4d top). We finally also used the implant itself in a benchtop setup (Fig. 4e) and wirelessly controlled its power up. We confirmed the behaviour in this case remained the same (Fig. 4d bottom).

Overall, we conclude that the prototype electronics will measure an ORP that is at an offset from the true value. This offset may vary with medium properties such as conductivity and flow, but *in vitro* measurements averaged ~0-70 mV for fresh/stirred media (see Fig. S6 for additional examples) and up to ~120 mV for multi-day old/unstirred fecal matrices – including custom electrode offsets. These errors are within the variability reported in ORP literature[2] and smaller than many ORP-associated phenotypes observed. For instance, a general purpose front end such as the present one could still be used to distinguish between conventional and streptomycin treated mice (~400 mV) [58], mono- vs. hexa-colonized mice (~225 mV) [59], conventional and germ-free rats (~300 mV) [35], and conventional and germ-free mice (~450 mV) [59]. Subtler phenotype differences could be measured by implementing a high input impedance buffer in the electronics front-end.

To derive intuition for the observed phenomena and guide the future design of improved electronics for this application, we implemented approximate electrical modelling of the electrode sensor using lumped elements. The transient responses we observed in the reported experiments in Fig. 4 tend to follow exponential decays and settling to a different voltage, implying the electrode electrical response is described by large RC time constants. Differences in measured ORP in stirred and unstirred media (example in Fig S6) suggest that mass transport of the redox active species in the fecal matter can also affect the electrode impedance. Notably, in test solutions (e.g. Zobell's, Light's), a 10 M$\Omega$ voltmeter will return the same value as a 1 T$\Omega$ meter suggesting that the electrode impedance in such media is lower (supplemental Fig. S7). The transient responses allow us to (i) predict their approximate final settled value using curve fitting, and to (ii) map the fitted responses to a simple circuit model similar to the Randle circuit (Fig. 4f). The model component values $R_1$, $R_2$, C can be derived based on the beginning and end voltages of the transient, as well as the measurement instrument resistance (see supp. Fig. S8). The accuracy of this model could be improved by taking into account 2nd order effects, such as the impact of medium diffusion (example curve fits using two exponential time constants are shown in Fig. S4). Analysis techniques such as Electrochemical Impedance Spectroscopy (EIS) could also be used in the future to model the electrode system in more detail.

## IV. IN VIVO DEMONSTRATION

To demonstrate the sensor's ability to enable extended and convenient measurement of gut ORP *in vivo*, we implanted a fully functional sensor in a 12-week-old female Sprague Dawley rat. A second rat of the same age, sex, and breed was housed separately, did not receive an implant and served as a control to investigate possible effects of implantation. Animal handling was in accordance with a research protocol approved by the Institutional Animal Care and Use Committee (IACUC).

The animal receiving the implant was anesthetized using inhalant isoflurane titrated to effect (1.5-3%, 1-3 L/min of 100% oxygen) and maintained on circulating warm water blanket during surgery. Pre-emptive analgesia was provided using buprenorphine SR (1 mg/kg SQ) and meloxicam (0.2 mg/kg SQ). Surgical instruments and the implant were autoclaved prior to the surgery. The abdomen was shaved and sterilized using 3x alternating swipes of alcohol and betadine. Laparotomy was then performed and the cecum was exteriorized, isolated using a sterile swab, and kept moist with sterile saline. A 1 cm incision was made in the tail of the cecum along greater curvature, ensuring that leakage of cecal contents was minimal (see Fig. 5a). The implant was gently placed at the wall of the cecum and fastened by sutures through the wings to the surrounding cecum epithelium (Fig. 5a and b), and then secured within the cecum with non-absorbable suture in a continuous inverting pattern to close the cecal incision. The cecum was thoroughly rinsed to reduce contamination, and then replaced into the abdomen. The laparotomy was closed in 2 layers. Finally, the rat was monitored until fully recovered. Notably, by using a sterile implant and meticulous sterile technique, we were able to avoid prophylactic antibiotic treatment of the implanted rat; this enables experiments in antibiotic naïve animals, which are critical for studying the microbiome. Similar technique may also enable implantation into germ-free animals, though the technical hurdles are larger.

The rat recovered under daily monitoring for 3 days until activity returned to a pre-surgery baseline. On day 4 post surgery, an ORP measurement was taken and measurements continued at a frequency of approximately once every 2 days until day 12. For all measurements, the rat was placed under inhalant isofluorane anaesthesia (1.5-3%, 1-3 L/min of 100% oxygen) and kept on a warm blanket as above. Ultrasound gel was placed on the abdomen area and the ultrasonic transducer was gently placed on the rat skin (Fig. 5c). The device location was identified through gentle massage of the rat abdomen. Measurement lasted under 15 minutes, allowing for assembling the measurement setup, locating the sensor, and taking sufficient samples; afterward the rat was allowed to recover.

Captured sensor codes (example shown in Fig. 5d) were mapped to ORP using the previously measured post-packaging calibration curve (Fig. 3c, in red). Similar transients were observed as in experiments in fecal matrices – as a result, the final values of the transients were used, as best estimates of the unloaded ORP (see supplemental Fig. S9 for example raw results). The measurement results are shown in Fig. 5e. Our ORP measured range was approx. -470 mV to -320 mV. Immediately post-euthanasia, using a Pt and standard Ag/AgCl reference (Fig. S10), the implanted rat cecum ORP measured -380 mV to -200 mV and the control rat measured -407 mV to -218 mV, suggesting the impact of the implant itself on cecal ORP was minimal. Values ranged widely and depended on location in cecum and proximity to the atmosphere (see supplement Fig. S11 for a demonstration of ORP change in fecal media transitioned from anaerobic to aerobic conditions). Overall, the combined error from the implant input impedance and custom electrode [Cl⁻] dependence which we discussed in previous sections is likely on the order of 100 mV. Nevertheless, literature values for ORP of exposed cecal content in sacrificed, anesthetized and ambulatory rats [34]–[36], and our own measurements broadly agree with our implant measured values. Note that *in vivo* ORP measurements in fully enclosed rat ceca have not been reported before, so no direct comparison is available.

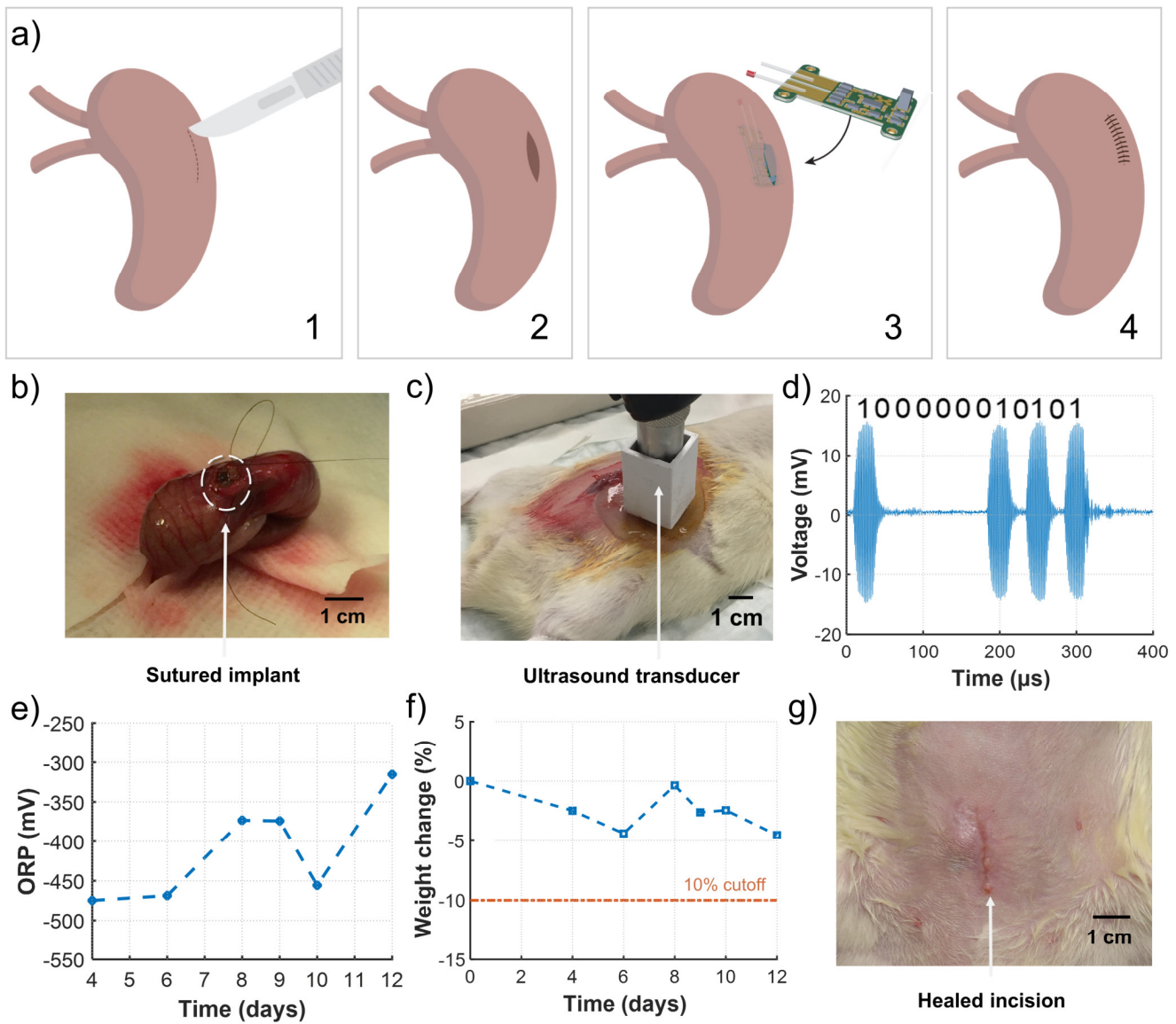

Fig. 5: *In vivo* procedures and results. a) Steps of implantation procedure: incision, opening of the cecum, placement of implant such that the electrodes point towards the head of the cecum without contacting the walls, and closure via suture. b) Sensor in the process of implantation and suturing in the rat cecum. c) Ultrasound measurement setup. d) Example of captured high SNR wireless signal from the sensor. e) Measured ORP from the implanted sensor over 12 days. f) Weight fluctuation throughout the experiment, remaining below the 10% humane endpoint. g) Healed incision 12 days after surgery, indicating surgical recovery and health. Note: scale bars shown in above photographs are approximate.

Throughout the 12-day experiment, the rat weight fluctuation remained well below the defined humane endpoint of 10%, as shown in Fig. 5f, indicating the implant was well tolerated. Some variations in weight can be attributed to different times of measurement relative to the animal feeding cycle. Inspection of the surgical incision at the last day of measurement (Fig. 5g) also indicated undisturbed healing. The cecum itself appeared healthy (Fig. S12) without gross changes in shape or size.

## V. Conclusion

We have developed an *in vivo* oxidation-reduction potential implant sensor measuring 20 mm x 5.6 mm x 4 mm and featuring wireless powerup and telemetry using ultrasound. The design and thorough characterization of electronics, packaging, and electrode materials enabled the prototype implant to survive standard autoclave sterilization and measure ORP for 12 days implanted in a rat cecum. This is the first implant able to produce longitudinal measurements of the ORP of a laboratory animal, enabling studies with alterations to diet, microbiome, and host genetics. While previous measurements of ORP have been conducted in laboratory animals, they have been almost exclusively terminal, preventing testing of most key experimental questions.

Our future work includes modifying the sensor electronics to produce a higher impedance measurement device and improving electrodes to reduce dependence on external [Cl⁻]. Combined, these changes should substantially lower the absolute error, allowing tests of subtle ORP phenotypes. Fabrication time can also be improved, allowing higher-n studies to measure the effect of diet, antibiotic, and microbiome composition on ORP.

Mechanistic insight into how ORP potentiates or ameliorates disease could lead to clinical translation by identifying patterns for diagnosis, and strategies for treatment, of diseases associated with redox imbalance (e.g. inflammatory bowel disease). While we believe ORP merits substantial investigation, the design and fabrication strategy of our implant is modular and could enable many other potentiometric electrochemical measurements (e.g. pH) in the gut.

## Acknowledgment

The authors would like to thank Joerg Deutzmann, Steven Higginbottom, Prof. Alfred Spormann, Ting Chia Chang, Kat Ng, Marios Sophocleous, Max Wang, and Ahmed Sawaby at Stanford University for their valuable input and helpful discussions. We also acknowledge Mentor Graphics for the use of the Analog FastSPICE (AFS) Platform and Stanford Nanofabrication Facility for the use of fabrication equipment.

# *In Vivo* Wireless Sensors for Gut Microbiome Redox Monitoring

Spyridon Baltsavias, Will Van Treuren, Marcus J. Weber, Jayant Charthad, Sam Baker, Justin L. Sonnenburg, and Amin Arbabian

S. Baltsavias, M. J. Weber, J. Charthad, and A. Arbabian are with the Electrical Engineering Department, Stanford University, Stanford, CA 94305 USA (contact email: sbaltsav@stanford). W. Van Treuren and J. L. Sonnenburg are with the Department of Microbiology and Immunology, Stanford University, Stanford, CA, 94305, USA. S. Baker is with the Department of Comparative Medicine, Stanford University, Stanford, CA 94305, USA

<div align="center">SUPPLEMENTARY INFORMATION</div>

### A. Materials

All chemicals used were acquired from Sigma-Aldrich, with the exception of: Zobell's, Modified Zobell's and Light's solutions obtained from Ricca Chemicals, PDMS Sylgard 184 from Dow Corning, Parylene C (Dichloro-p-cyclophane) from Specialty Coating Systems, nonconductive epoxy (FDA 2) from AA Bond, and conductive epoxy (Silver Epoxy) from MG Chemicals. For ORP vs. standard reference measurements in Fig. 3a and Fig. 3d, we used a SympHony SB70P pH meter in mV mode and a double junction Ag/AgCl reference electrode from Sigma-Aldrich. For the electrode characterization measurements in Fig. 3b, we used PCE-228 ORP loggers and Ag/AgCl reference electrodes (REDOX ORP-14, PCE Instruments). For ORP measurements in Fig. 4a-d we similarly used the PCE-228 ORP loggers; the traces labeled Pt vs. Std. ref. use the double junction Sigma-Aldrich Ag/AgCl electrode as reference. Platinum wires were acquired from SurepureChemetals (item 6044). The solid-state reference electrodes (Refex) were fabricated at Refex Sensors Ltd. A Labcoter 2 PDS 2010 was used for Parylene C coating. All ultrasound measurements used the Olympus A303S transducer, Keysight 33500B Waveform generator and E&I Model 411LA power amplifier. Finally the Agilent Precision Impedance Analyzer 4294A was used for piezo impedance characterization and the Keysight B2962A for electronics resistance characterization.

The pseudo-reference electrodes used in Fig. 3b were made using Dropsens C220AT screen printed Silver electrodes manually coated with an Ag/AgCl paste (ALS Co., Ltd). The paste was cured at 65 °C for 1 hour. PBS for the experiment also shown in Fig. 3b was prepared by diluting 10X stock to 1X with distilled water for a final concentration of 0.137 M NaCl, 0.0027 M KCl, and 0.0119 M phosphates.

### B. Ultrasound transmission and data recovery

Measurements involving implant ultrasound power up and data recovery used the A303S ultrasonic transducer with a focal distance of 3 cm. To maintain the implant piezo within the focal zone, a custom 2 cm long coupler filled with ultrasound gel was attached to the transducer resulting in a total transmission distance of 3 cm. To transmit ultrasound, we used the waveform generator, amplified by the power amplifier. Based on previous characterization of the transmitter's beam profile, we estimate the peak acoustic intensity at the piezo at this distance to be 8.6 mW/mm$^2$. For a repetition rate of 1 kHz, and a duty cycle of 12.7 % to 15.2 % (100 to 120 cycles of 790 kHz carrier), the time-averaged acoustic intensity is $\leq$ 1.3 mW/mm$^2$ which is almost 6x below the FDA limit. Given our piezoelectric material area (1.08 mm x 1.08 mm), aperture efficiency (0.2 measured with a fully encapsulated device at 790 kHz), and above input acoustic intensity during charging, we calculate an available electrical power of 2 mW and an available energy of 278.5 nJ each power-up repetition (assuming 110 cycles of 790 kHz ultrasound carrier). The implant typically uses 131 nJ to operate, giving a power-up energy efficiency of 47%.

A custom clipper diode-based transmit/receive switch was used to receive ultrasound signals from the implant after the end of power transmission. The captured data from the transducer were displayed and saved using the oscilloscope. The uplink data is encoded in OOK and consists of 10 bits. To interpret this data we performed OOK demodulation in Matlab using envelope detection and thresholding methods. The result is a code from 0 to 1023 ($2^{10}$-1), expected to map to an ORP voltage range of approximately -500 mV to 300 mV.

*C. Supplementary figures*

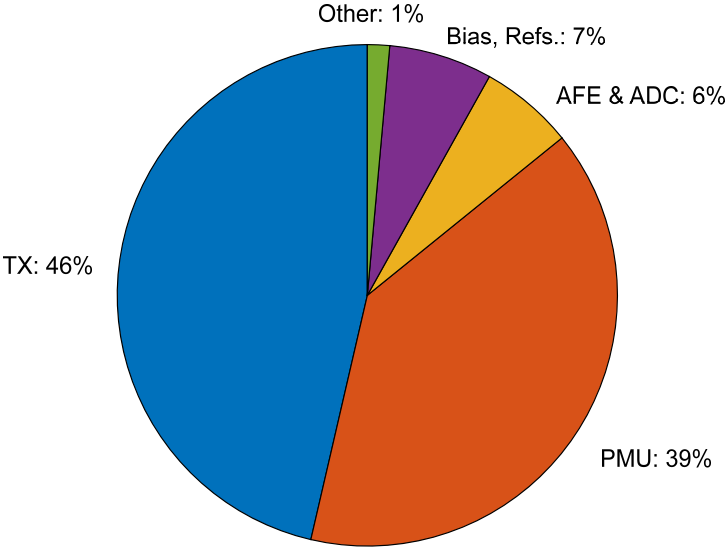

Fig. S1: Implant average power consumption breakdown for a transmitted code 1111111111. The total average power was 152.5 µW from simulation. Since OOK modulation is used where the transmitter (TX) is turned off during bit 0, this represents the maximum power dissipation.

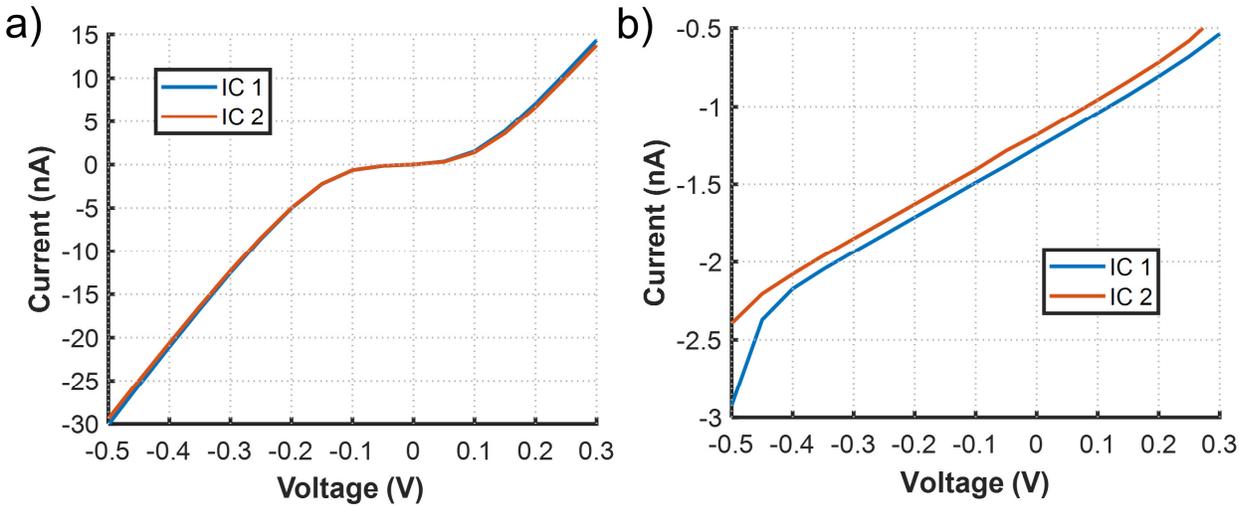

Fig. S2: Implant electronics front-end resistance characterization. a) Current-voltage plot for two implant ICs in powered off mode. b) Current-voltage plot for two implant ICs in powered on mode. In both cases the worst-case leakage current is observed for an input of -0.5 V. Our tests therefore use an equivalent resistance at this voltage value to simulate the effect of the chip on electrodes and ORP. Measurement details: a Keysight B2962A connected to the input leads of the implant was used to sweep the voltage and measure the current. Shielded insulated cables were used to minimize noise and cable leakage. At each 50 mV voltage increment, 100 measurements of the current were averaged together.

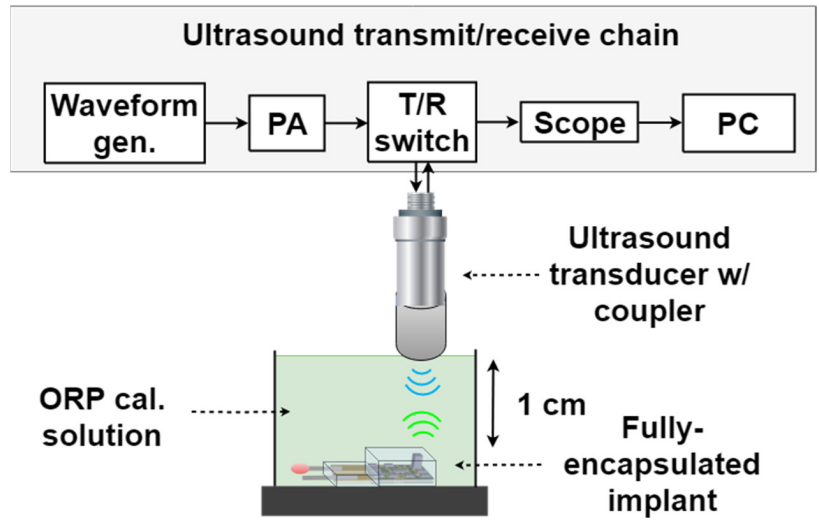

Fig. S3: Conceptual diagram of the *in vitro* electronics measurement setup, after sensor encapsulation. The implant was powered and interrogated similarly to Fig. 3f (details in "Ultrasound transmission and data recovery" section). In this case the encapsulated implant was immersed directly in the calibration solution of interest. To compare to the sensor readout, a pair of standard ORP electrodes were also immersed in the medium and connected to the ORP voltmeter (not shown here).

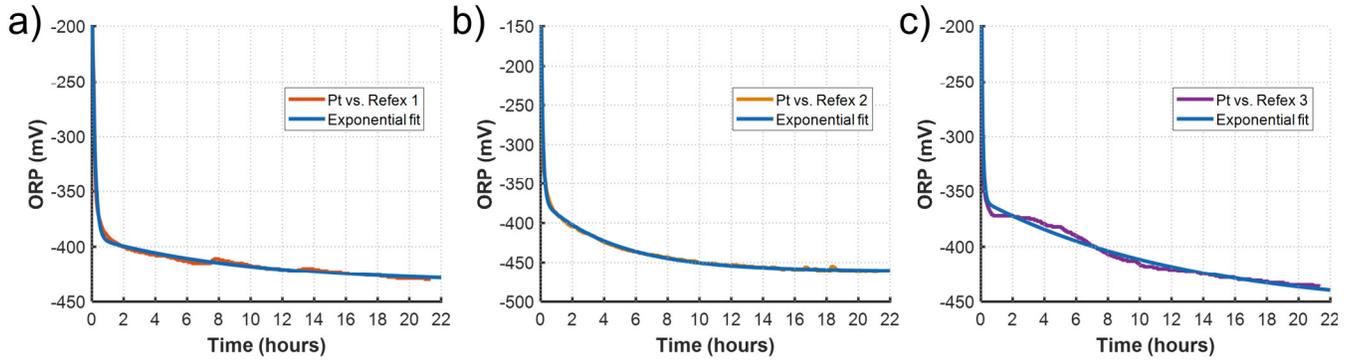

Fig. S4: Estimation of exponential asymptote in ORP transients. The above traces correspond to the end of the experiments shown in Fig. 4a, where t = 0 is the switch from low impedance loading (17 MΩ) to high impedance (1 TΩ). The Matlab curve fitting tool (cftool) was used to compute the above fits given the equation $f(t) = a * e^{-bt} + c * e^{-dt} + e$. For t → ∞, $f(t) = e$ and thus this term was used to estimate the settled values. a) $e = -432.7$, fit $R^2$: 0.9865. b) $e = -461.3$, fit $R^2$: 0.9916. c) $e = -458.2$, fit $R^2$: 0.9698.

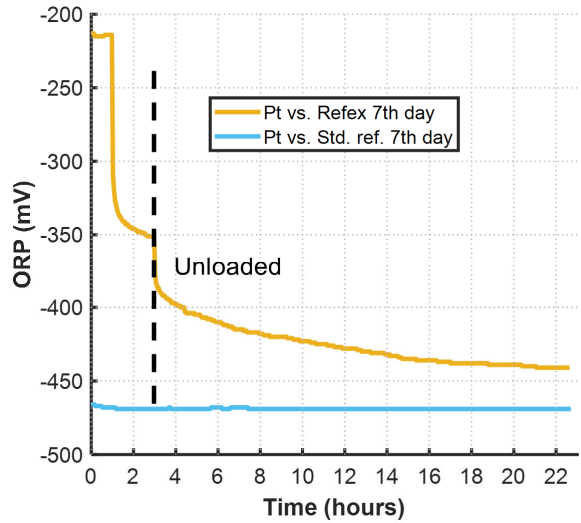

Fig. S5: The extended last transition of the 8-day experiment of fecal immersion (Fig. 4c of the main text). The difference between the unloaded measurement of the custom electrode pair is seen to decrease over time. At 22 hours the electrode recovers to within 25 mV of the ground truth measurement.

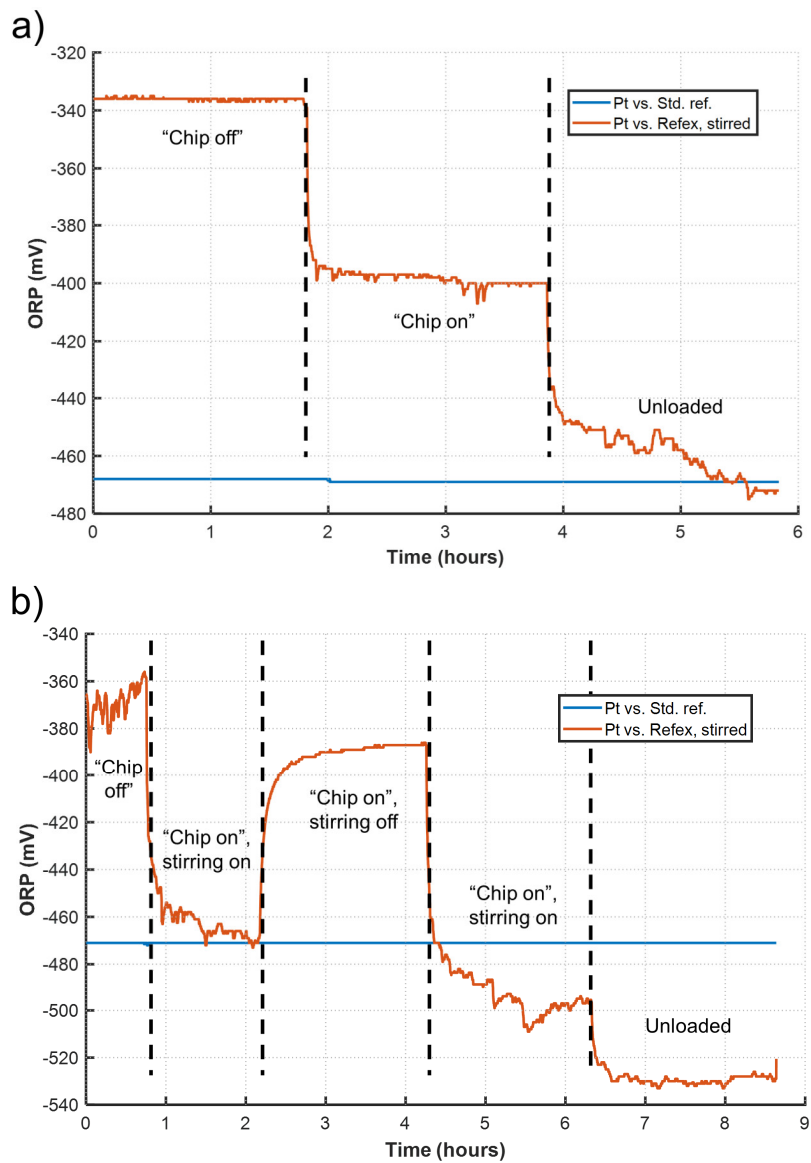

Fig. S6: a) Example of impedance effect in stirred fecal media. Similar behavior is seen in unstirred media. Here the electronics on-mode was approximately 70 mV higher than the ground truth (standard reference measurement). b) Stirring is observed to result in a significantly different ORP, with values closer to the ground truth compared to an unstirred medium; this indicates electrode impedance is influenced by medium convection and flow.

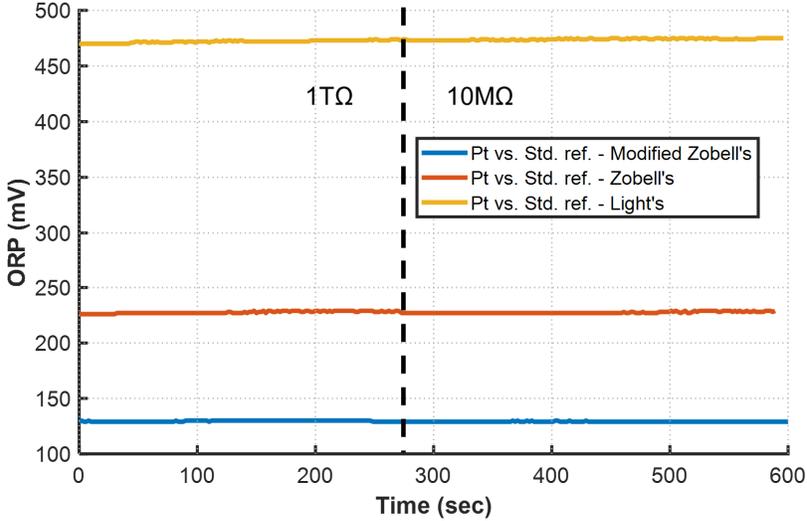

Fig. S7: Three electrodes in ORP calibration media are measured with a PCE-228 1 TΩ logger and then loaded with a 10 MΩ resistor. The recorded ORP does not change, indicating the electrode impedance is significantly lower than in feces.

a)
b)

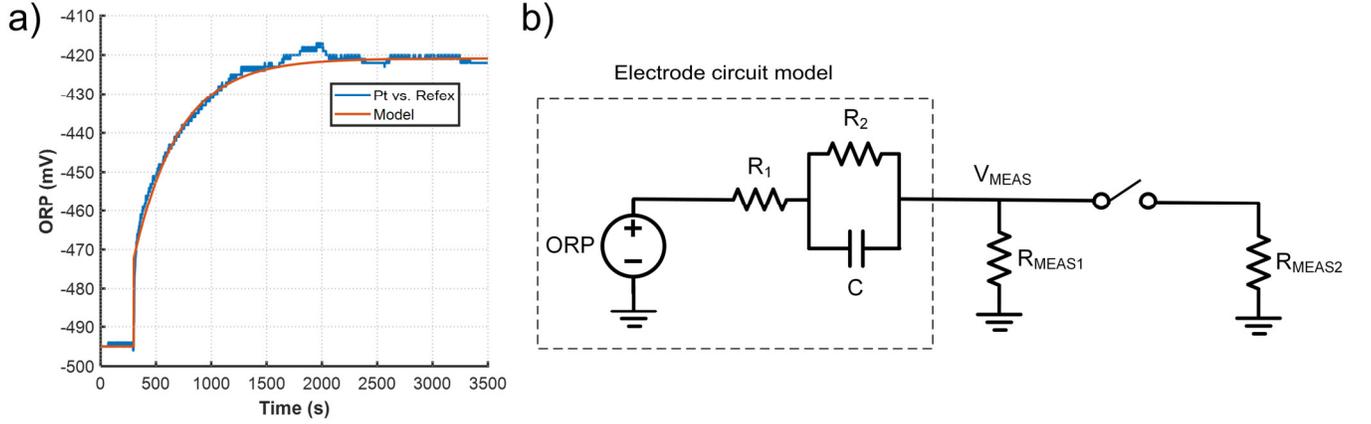

Fig. S8: Approximate modeling of electrode transient response due to a step in connected resistance. In a) (also Fig. 4f) a transition from a 1 TΩ measurement resistance to a 17 MΩ resistance connected in parallel is shown. While not all transitions we have measured follow the same response (e.g. Fig. S4), the simple circuit in b) can often model the electrode system and measurement resistances well. Here $R_{MEAS1}$ = 1 TΩ, $R_{MEAS2}$ = 17 MΩ and ORP, $R_1$, $R_2$ and C are electrode and medium related impedance parameters. Wire resistance/capacitance and other measurement/instrument capacitances were assumed to be negligible.

The response of the circuit in b) can be expressed as:

$$V_{MEAS}(t) = \begin{cases} V_{MEAS,t0^-} & for\ t < 0\ (before\ switch) \\ (V_{MEAS,t0^+} - V_{MEAS,t\infty}) \cdot e^{-\frac{t}{\tau}} + V_{MEAS,t\infty} & for\ t > 0\ (after\ switch) \end{cases},$$

where $V_{MEAS,t0^-}$, $V_{MEAS,t0^+}$ and $V_{MEAS,t\infty}$ correspond to the measured voltage before the switch closes, the measured voltage right after the switch closes and the final settled voltage respectively, and τ is the system time constant. These are parameters that can be extracted directly from the experimentally measured response by curve fitting. In the simplified case of $R_{MEAS1} >> R_1 + R_2$ and $R_{MEAS1} >> R_{MEAS2}$, such as the one shown, they can be then mapped to the electrode and medium related parameters as follows:

$$V_{MEAS,t0^-} = ORP$$

$$V_{MEAS,t0^+} = ORP \cdot \frac{R_{MEAS2}}{R_{MEAS2} + R_1}$$

$$V_{MEAS,t\infty} = ORP \cdot \frac{R_{MEAS2}}{R_{MEAS2} + R_1 + R_2}$$

$$\tau = C \cdot (R_{MEAS2} + R_1) || R_2 = C \cdot \frac{(R_{MEAS2} + R_1) \cdot R_2}{R_{MEAS2} + R_1 + R_2}$$

Solving for the model unknowns in this case we get ORP = -495 mV, $R_1$ = 828 kΩ, $R_2$ = 2.17 MΩ, C = 213 µF approximately.

In cases where $V_{MEAS,t0^-} \approx V_{MEAS,t0^+}$, the model can be simplified by taking $R_1 = 0$.

This analysis could be generalized for arbitrary $R_{MEAS1}$ and $R_{MEAS2}$, and more complicated models could be developed for responses better fitted by two exponential time constants such as in Fig. S4.

Note that while this process can be used to derive intuition for resistance transitions, despite the resemblance of the model to the Randle circuit, the derived parameters may not directly map to physical properties of the electrode and medium interface, such as the double layer capacitance and faradaic resistance.

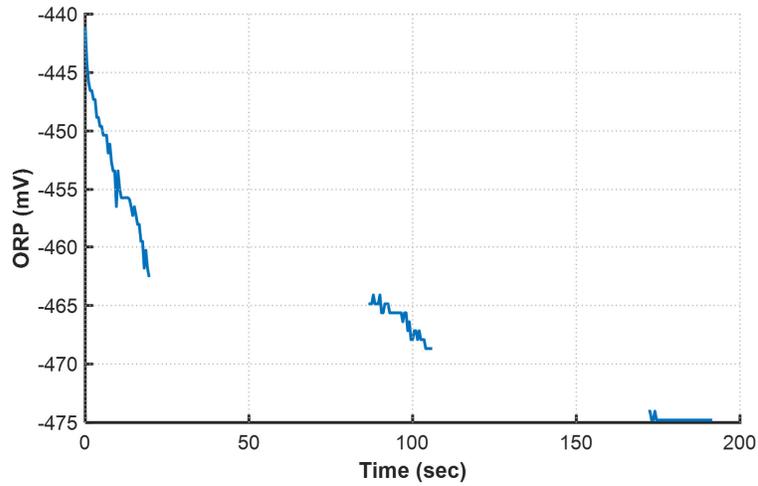

Fig. S9: Sample *in vivo* data vs. time. The raw binary data were demodulated to voltage using the post-encapsulation calibration curve shown in Fig. 3d. A Matlab script was used to log data of given length; due to the limitations of this script there are gaps in the data during which the data is saved. Nevertheless the data is seen to follow transient responses similar to those in *in vitro* tests due to limited input resistance.

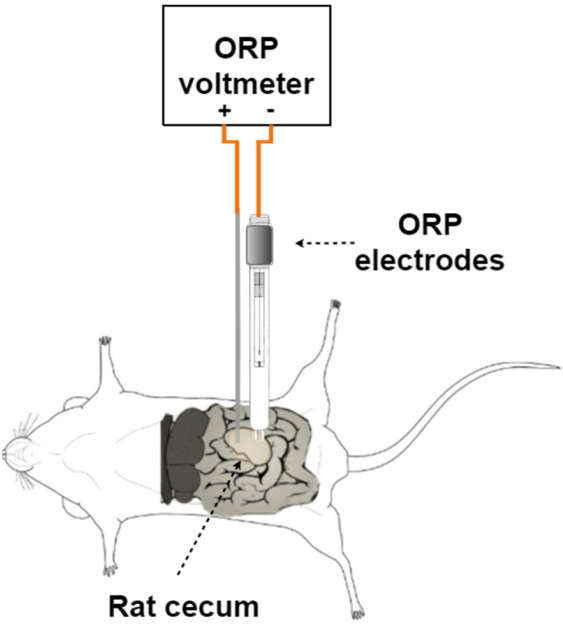

Fig. S10: Conceptual diagram of post-euthanasia ORP measurement. The standard ORP electrodes (platinum and Sigma Aldrich Ag/AgCl reference) are inserted in the incised and exposed rat cecum.

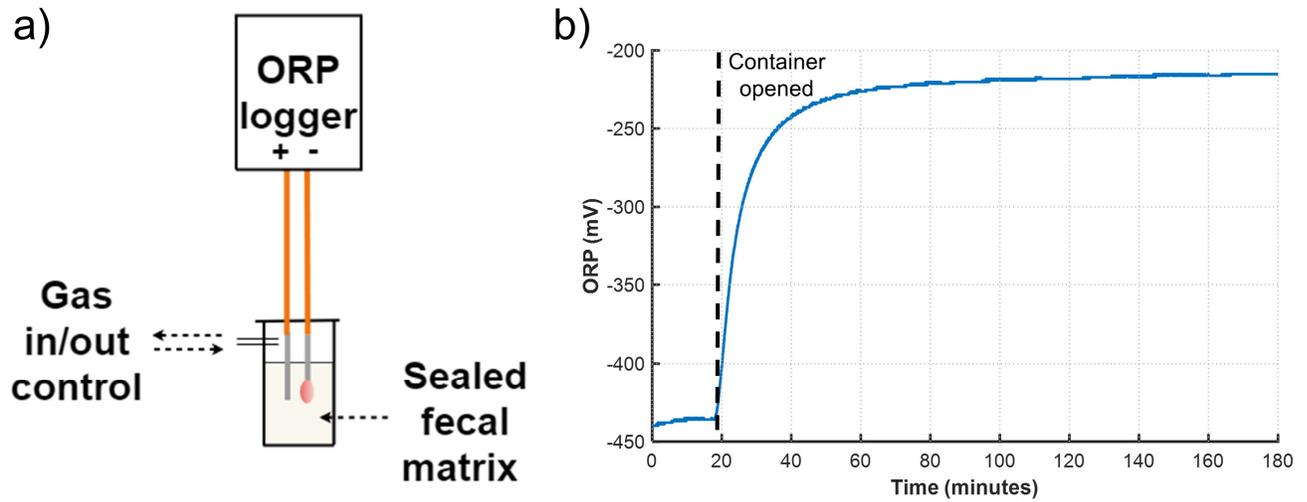

Fig. S11: Demonstration of the effect of atmospheric changes on ORP. a) A fecal slurry was incubated in an anaerobic chamber for 14 days and subsequently sealed in a container with ORP electrodes immersed in the medium, similarly to Fig. 4e. The tube was then moved to an aerobic environment and an opening was made, while the ORP was recorded. b) The results indicate that exchange of gas between media and atmosphere – likely loss of hydrogen from the media and ingress of oxygen – produces an increase in ORP.

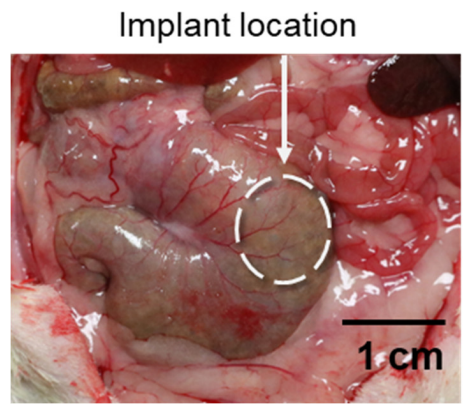

Fig. S12: Implant location in the rat cecum. The photo was taken during necropsy and indicates the rat intestines are healthy. Shown scale is approximate.